# Laser spectroscopy of the $X\ ^1\Sigma^+$ and $B\ ^1\Pi$ states of the LiRb molecule

Sourav Dutta[a,*], Adeel Altaf[a], D. S. Elliott[a,b,†] and Yong P. Chen[a,b,‡]

[a] *Department of Physics, Purdue University, West Lafayette, IN 47907, USA*
[b] *School of Electrical and Computer Engineering, Purdue University, West Lafayette, IN 47907, USA*
[*] *sourav.dutta.mr@gmail.com,* [†] *elliottd@purdue.edu,* [‡] *yongchen@purdue.edu*

We have studied the $X\ ^1\Sigma^+$ and $B\ ^1\Pi$ states of $^7$Li$^{85}$Rb using Laser Induced Fluorescence (LIF) spectroscopy and Fluorescence Excitation Spectroscopy (FES). We extract molecular constants for levels $v'' = 0$–$2$ of the $X\ ^1\Sigma^+$ state and levels $v' = 0$–$20$ of the $B\ ^1\Pi$ state. For the $B\ ^1\Pi$ state, we have observed rotational perturbations in the $e$–parity component of the $v' = 2$ level, and determined the dissociation energy. We discuss implications of our measurements in finding efficient photoassociation pathways for production of ultra-cold ground state LiRb molecules, and their detection via state selective ionization.

## 1. Introduction

The spectroscopy of heteronuclear diatomic bi-alkali molecules provides critical information for the rapidly developing field of ultra-cold molecules [1–4]. The thermal motion of molecules in an ultra-cold sample is much slower than a room temperature gas, reducing collisions between molecules significantly. In this limit of reduced collisions, quantum mechanical effects are prominently visible; and manipulating chemical reactions by controlling molecular collisions becomes possible [1–4]. There has been significant effort and progress in creating ultra-cold deeply bound ground state bi-alkali diatomic molecules [2,3], such as Cs$_2$, K$_2$, Rb$_2$, KRb, NaCs, LiCs etc. Among these, the heteronuclear species are of great interest due to their relatively large permanent electric dipole moment [5], with LiRb predicted to have an electric dipole moment of 4.1 Debye. These heteronuclear molecules provide a good system not only for studying ultra-cold chemistry [1–4], but also for studying quantum phase transitions [3,6] and implementation of various quantum computing protocols [3,7]. Recently two groups reported observation of Feshbach resonances in ultra-cold mixture of Li and Rb atoms [8,9]. These studies provide good estimates of the scattering length, but efficient formation and detection of ultracold LiRb molecules also requires detailed knowledge of bound rovibrational levels of the ground and excited states.

Until the very recent work of Ivanova *et al.* [10], the LiRb molecule was the only heteronuclear diatomic bi-alkali molecule for which no detailed spectroscopic investigation had been reported. In ref. [10], the ground $X\ ^1\Sigma^+$ state was studied using Fourier transform LIF spectroscopy and the potential energy curve was constructed. There have been no reports on experimental studies of the excited electronic states of this molecule. The first theoretical calculation of the potential energy curves for LiRb was carried out by Igel-Mann *et al.* [11] in 1986. More recently, Korek *et al.* [12,13] calculated the *ab initio* potential energy curves for LiRb – some of which are shown in Figure 1. The excited $B\ ^1\Pi$ state of LiRb is a promising intermediate state for the formation of ultra-cold LiRb molecules by photoassociation [14,15], and also for their detection via state selective ionization. In this article we report, for the first time, a high resolution spectroscopic study of the $B\ ^1\Pi$ state using Fluorescence Excitation Spectroscopy (FES). In addition, we also briefly report an independent study of the $X\ ^1\Sigma^+$ state.

## 2. Experiment

We performed the measurements on LiRb molecules produced in a 80 cm long three-section heat-pipe oven similar to the one described by Bednarska *et al.* [16]. This type of heat pipe allows us to produce Li and Rb vapors at similar densities, despite the large vapor pressure difference of these metals at a common temperature. We loaded 10 g of Li (natural isotopic abundance) in the central section (~15 cm long) of the heat pipe oven, which was heated to 550 ˚C. The two outer sections (each ~10 cm long), each containing 5 g of Rb (also natural isotopic abundance), were maintained at a lower temperature of 300 ˚C. We used Argon as a buffer gas at a pressure of approximately 4–5 Torr and water cooled the two ends of the heat-pipe oven to protect the optical windows from deposition of metal. Under these conditions, LiRb molecules formed in the central section of the heat-pipe oven (along with Li$_2$ and Rb$_2$ molecules). We have operated the heat-pipe oven for more than 800 hours over a period of 14 months without refilling.

The LiRb molecules formed in the heat pipe are distributed among different vibrational and rotational levels of the ground electronic state ($X\ ^1\Sigma^+$ state). We excite these $X\ ^1\Sigma^+$ state molecules to the $B\ ^1\Pi$ state with the single mode output of a cw, frequency stabilized, ring dye laser. The ring dye laser (with Rhodamine 6G dye) is pumped with a 6 W Verdi laser (532 nm) and operates at frequencies in the range 16300-18025 cm$^{-1}$. As shown in Figure 2A, we collected the laser induced fluorescence from one end of the heat pipe oven and directed it toward a 0.75 m focal length Czerny-Turner monochromator (SPEX 1702) with a typical resolution of 3 cm$^{-1}$. The light is detected at the exit slit with a photomultiplier tube module (Hamamatsu H9306-03). We have observed transitions in $^7$Li$^{85}$Rb and $^7$Li$^{87}$Rb, but only $^7$Li$^{85}$Rb will be considered in this article. We did not attempt to record transitions from $^6$Li$^{85}$Rb and $^6$Li$^{85}$Rb and did not observe any lines originating from Rb$_2$ or Li$_2$ in this frequency range.

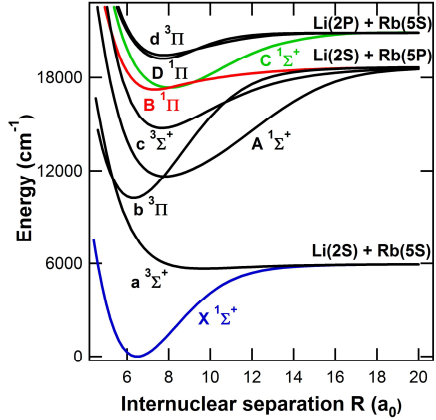

**Figure 1.** LiRb potential energy curves from theoretical *ab initio* calculation of Korek *et al.* [12]

We performed two types of measurements: Laser Induced Fluorescence (LIF) spectroscopy and Fluorescence Excitation Spectroscopy (FES). In the traditional LIF measurements, we kept the laser frequency fixed while scanning the monochromator and recorded the rotationally-resolved LIF spectrum (Figures 2B, 2C). The LIF spectrum directly gives an estimate of the vibrational spacing for $X\ ^1\Sigma^+$ state of $^7$Li$^{85}$Rb (~ 193 cm$^{-1}$). For the FES measurements, we use the scheme shown in Figure 3A. We fix the monochromator to detect fluorescence at a frequency ~ 193 cm$^{-1}$ less than that of the excitation frequency, and scan the dye laser frequency to record a spectrum such as the one shown in Figure 3B. Peaks in these spectra correspond to excitation of specific ro-vibrational levels of the $B\ ^1\Pi$ state. The average accuracy of the line positions in FES is 0.02 cm$^{-1}$, limited by the accuracy of the wavelength-meter used for measuring the laser wavelength. We used the higher resolution FES measurements for determination of the molecular constants of both the $X\ ^1\Sigma^+$ and $B\ ^1\Pi$ states. In the following, we briefly discuss the measurements performed with LIF spectroscopy, followed by a more detailed account of the higher resolution FES measurements.

### 3. Description and analysis

We used the LIF spectra to extract useful information about the vibrational and rotational structure of the $X\ ^1\Sigma^+$ state. As shown in Figure 2B, we were able to record LIF up to $v'' = 45$ of the $X\ ^1\Sigma^+$ state (covering more than 98% of the potential well depth). We observed a few weak lines beyond $v'' = 45$ that do not fit the LIF progression of the $X\ ^1\Sigma^+$ state. This might indicate LIF to the $a\ ^3\Sigma^+$ state, although more work is needed to confirm this. We have used our LIF measurements to determine the molecular constants of the $X\ ^1\Sigma^+$ state, and find excellent agreement with the Fourier transform spectroscopy measurements of Ivanova *et al.* [10]. In light of this strong agreement, and since the Fourier transform results were of higher spectral resolution than our LIF measurements, we will not repeat these results here. Instead, we will focus mainly on the excited $B\ ^1\Pi$ state, which has not been studied before.

The intensity variations in the LIF spectra provide a good picture of the nodal structure in the vibrational wavefunction of the excited state, and we use these to assign the vibrational number $v'$ of the excited $B\ ^1\Pi$ state. In Figure 2C we show the LIF spectra obtained upon excitation of four different vibrational levels of the $B\ ^1\Pi$ state. As expected from the Franck-Condon principle, the intensity distribution of LIF from $v' = 0$ has no nodes, $v' = 1$ has one node, $v' = 2$ has two nodes and so on.

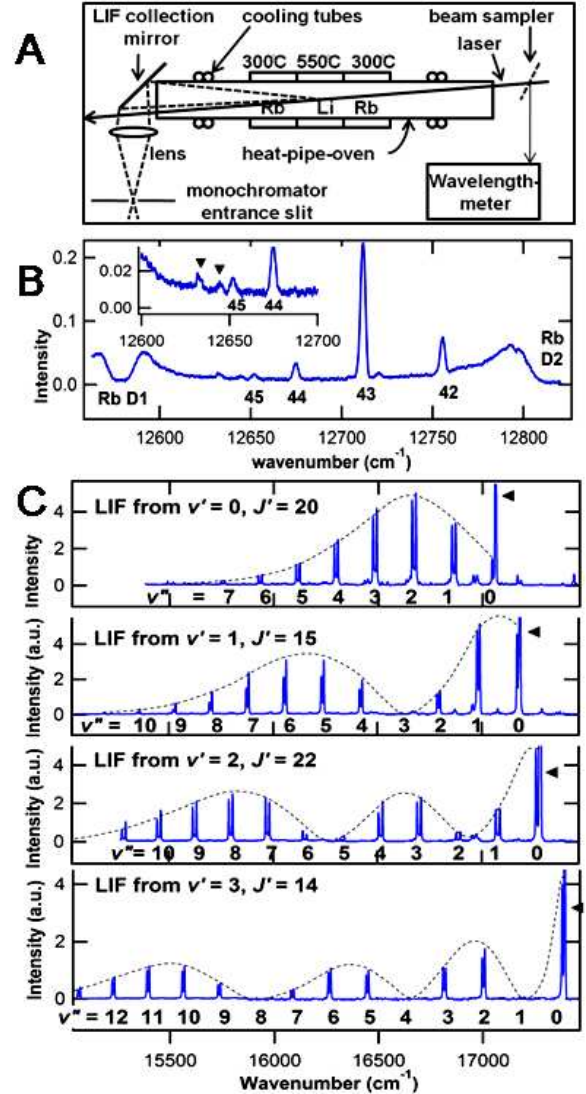

**Figure 2.** (A) Experimental setup: LiRb molecules formed at the center of the heat-pipe-oven are excited with a laser, the LIF is collected in the forward direction and sent to a monochromator. (B) LIF after exciting the transition $(v'' = 2, J'' = 6) \rightarrow (v' = 20, J' = 6)$ with the laser frequency at 18016.42 cm$^{-1}$. The LIF progression can be followed up to $v'' = 45$. The features due to self absorption of Rubidium vapor are also seen. Inset: Zoomed in view showing the highest observed $v''$ levels of the $X\ ^1\Sigma^+$ state along with two unassigned transitions (indicated by ▼). (C) LIF spectra originating from $v' = 0, 1, 2$ and 3 levels of the $B\ ^1\Pi$ state and terminating in different $v''$ levels of the $X\ ^1\Sigma^+$ state. The intensity distribution clearly reflects the node structure of the radial wave functions of the $B\ ^1\Pi$ state. The occurrence of *P-R* doublets facilitates assignment of *J*. The dotted lines serve to guide the eye only. The line arising from scattered light at the laser frequency is indicated by ◄.

Once these assignments are made, we turn to FES for higher resolution measurements of $B\ ^1\Pi$ states. We used two modes for recording these measurements: complete spectra while tuning the laser frequency continuously in an approximately linear scan; or tuning the laser manually and recording the wavelengths and intensities of the peaks. In either case, we determine the wavelength of the laser output using a wavelength-meter (Burleigh WA-1000). In Figure 3B we show an example of the Fluorescence Excitation Spectra around the head of the ($v''=0$, $v'=2$) band, obtained by patching together several frequency scans (~ 13 GHz each). Assignments of several rotational lines belonging to the $P$, $Q$ and $R$ branches are indicated. We recorded similar high resolution FES for 27 different ($v''$, $v'$) bands of the $X\ ^1\Sigma^+ \to B\ ^1\Pi$ transition, where $v''$ ranged from 0–2 and $v'$ ranged from 0–20. For each of these vibrational bands we observed the $Q$ branch rotational levels and assigned lines up to $J \sim 20$. We did not attempt to study the higher $J$ levels because our ultimate goal is to study the formation of ultra-cold molecules via photoassociation, where only $J < 4$ levels are generally observed. We assigned the $P$ and $R$ branches ($e$ parity levels [17]) only up to $v' = 3$. We show plots of all observed $Q$ branch transition frequencies, against $J(J+1)$ in Figure S1 of the supporting material. A list of all observed transitions is also available in Table S1 of the supporting material. The analysis which follows is based on these high resolution FES measurements.

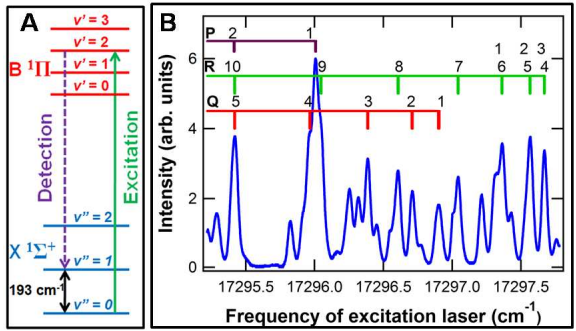

**Figure 3. (A)** Detection scheme for FES: The monochromator is fixed to detect fluorescence at a frequency ~ 193 cm$^{-1}$ less than that of the excitation, and the laser frequency is scanned. **(B)** A fragment of the Fluorescence Excitation Spectrum near the ($v'' = 0$, $v' = 2$) band head. The rotational quantum numbers $J'$ of the $P$, $Q$, and $R$ branches are assigned, as indicated on the top. At our resolution, levels $R(1)$, $R(2)$ and $R(3)$ overlap with $R(6)$, $R(5)$ and $R(4)$ respectively.

We determine the molecular constants in a level-by-level manner rather than using a global fit. The total energy $E_{v,J}$ of a given ($v$, $J$) level is expressed in the form [18]:

$$E_{v,J} = T_v + F_v(J). \quad (1)$$

$F_v(J)$ is the rotational energy and $T_v$ is the rotation-less energy for the vibrational state $v$,

$$T_v = T_e + G_v \quad (2)$$

where $T_e$ is the minimum of the corresponding electronic potential energy curve, and $G_v$ is the vibrational energy. As is commonly done [18], we approximate the value of the vibrational energy $G_v$ as an anharmonic oscillator and rotational energy $F_v(J)$ as a vibrating rotor:

$$G_v = \omega_e (v + \tfrac{1}{2}) - \omega_e x_e (v + \tfrac{1}{2})^2 + \omega_e y_e (v + \tfrac{1}{2})^3 + \ldots \quad (3)$$

$$F_v(J) = B_v [J(J+1) - \Lambda^2] - D_v [J(J+1) - \Lambda^2]^2 \quad (4)$$

where $\Lambda^2$ is 0 for the $X\ ^1\Sigma^+$ state and 1 for the $B\ ^1\Pi$ state, $B_v$ is the rotational constant, $D_v$ is the centrifugal distortion constant, $\omega_e$ is the harmonic constant while $\omega_e x_e$, $\omega_e y_e$ etc. are the anharmonic constants. Also, to a first approximation, the rotational constant $B_v$ is given by

$$B_v = B_e - \alpha_e (v + \tfrac{1}{2}) \quad (5)$$

For each vibrational level $v''$ from 0 to 2 of the $X\ ^1\Sigma^+$ state, we measure the difference in energy of the $P$ and $R$ lines using FES:

$$\Delta_2 F_{v''}(J'') \equiv E_{v'',J''+1} - E_{v'',J''-1}$$
$$= (4J''+2)B_{v''} - 2(J''^2 + J'' + 1)(4J''+2)D_{v''} \quad (6)$$

We determine the constants $B_{v''}$ and $D_{v''}$ from a least squares fit of $\Delta_2 F_{v''}(J'')$ vs. $J''$ (Figure 4A) and list the values of $B_{v''}$ in Table 1. We obtain $B_{e''} = 0.21661(7)$ and $\alpha_{e''} = 0.001575(4)$, from a fit of the data to eq. (5); these values are in good agreement with the higher precision results of Ivanova et al. [10]. The uncertainties in the values of $B_{v''}$ obtained from the fit are <0.025%. The relative uncertainties of the values of $D_{v''}$ is much larger, ~ 25%, due in part to the focus of our measurements on low-$J$ states. In light of this uncertainty, we neglect the variation of $D_{v''}$ with $v''$, and we report only an average value $D_{e''} = 1.26(10) \times 10^{-6}$ cm$^{-1}$. This value is also in reasonable agreement with $D_{e''} = 1.062 \times 10^{-6}$ cm$^{-1}$ reported in ref. [10], as well as the Kratzer relation for the Morse potential approximation: $D_{e''} = 4B_{e''}^3/\omega_{e''}^2 = 1.065 \times 10^{-6}$ cm$^{-1}$, using $B_{e''}$ and $\omega_{e''}$ from ref. [10]. For determination of all the other constants for the $X\ ^1\Sigma^+$ and $B\ ^1\Pi$ states, we restrict our analysis to the $Q$ branch transitions only.

The energy $Q(J)$ of an absorption line corresponding to a $X\ ^1\Sigma^+\ (v'',J) \to B\ ^1\Pi\ (v',J)$ transition is

$$Q(J) = E_{v',J} - E_{v'',J}$$
$$= (T_{v'} - T_{v''}) + B_{v'} [J(J+1) - 1] - D_{v'} [J(J+1) - 1]^2$$
$$- B_{v''} J(J+1) + D_{v''} J^2(J+1)^2 \quad (7)$$

With the values of $B_{v''}$ held fixed at the values reported in Table I and $D_{v''}$ held fixed at $1.06 \times 10^{-6}$ cm$^{-1}$, we obtain the values of $B_{v'}$, $D_{v'}$ and $(T_{v'} - T_{v''})$ from a least squares fit of $Q(J)$ vs. $J$ (Figure 4B). We obtain $T_{v''=1} - T_{v''=0}$ from the difference in frequencies between the following rotation-less transitions: $X\ ^1\Sigma^+\ (v''=1, J=0) \to B\ ^1\Pi\ (v', J=0)$ and $X\ ^1\Sigma^+\ (v''=0, J=0) \to B\ ^1\Pi\ (v', J=0)$. $T_{v''=2} - T_{v''=0}$ is calculated in a similar manner. $T_{v'}$ is then evaluated from the $(T_{v'} - T_{v''})$ values obtained above. In Table I, we report the values of $T_v$, $B_v$ and $D_v$ for both $X\ ^1\Sigma^+$ and $B\ ^1\Pi$ states.

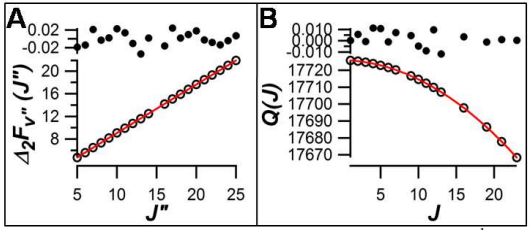

**Figure 4.** (A) A plot (open circles ○) of $\Delta_2 F_{v''}(J'')$, in cm$^{-1}$, for the $v'' = 0$ level of the $X\,^1\Sigma^+$ state. The solid line (red) is a fit to eq. (6) and the filled circles (●) are the fit residuals. $B_{v''}$ and $D_{v''}$ for the $v'' = 0$ level are obtained from this plot. The excited state involved was the $v' = 2$ level of the $B\,^1\Pi$ state. (B) A plot (open circles ○) of $Q(J)$, in cm$^{-1}$, for transitions from $v'' = 2$ to $v' = 13$. We obtain the values of $B_{v'=13}$, $D_{v'=13}$ and $(T_{v'=13} - T_{v''=2})$ from this plot. Similar plots are made for the Q-lines of other $(v'', v')$ transitions and the respective $B_{v'}$, $D_{v'}$ and $(T_{v'} - T_{v''})$ are obtained.

The energy differences $T_{v''=1} - T_{v''=0}$ and $T_{v''=2} - T_{v''=1}$ that we measure are in excellent agreement with those derived using the spectroscopic constants reported by Ivanova et al. [10]. From a fit of $T_{v'}$ to equations (2–3), we extract: $T_{e'}(B\,^1\Pi) = 17110.0(5)$ cm$^{-1}$ and $\omega_{e'} = 122.3(3)$ cm$^{-1}$. Corresponding theoretical values from Korek et al. [12] are $T_{e'}(B\,^1\Pi) = 17205$ cm$^{-1}$ and $\omega_{e'} = 140.5$ cm$^{-1}$. We do not make explicit comparisons to ref. [13] calculations, which included spin-orbit effects, because these results are expected to be more precise only for larger internuclear spacings.

We calculate the dissociation energy $D_e(B\,^1\Pi)$ of the $B\,^1\Pi$ state using the standard relation:

$$D_e(B\,^1\Pi) = D_e(X\,^1\Sigma^+) + \Delta E[Rb(5\,^2P_{3/2}) - Rb(5\,^2S_{1/2})] - T_{e'}(B\,^1\Pi)$$

where $D_e(X\,^1\Sigma^+) = 5927.9\,(40)$ cm$^{-1}$ is the dissociation energy of the $X\,^1\Sigma^+$ state as reported in ref. [10], $\Delta E\,[Rb(5\,^2P_{3/2}) - Rb(5\,^2S_{1/2})] = 12816.6$ cm$^{-1}$ is known very precisely from standard databases and $T_{e'}(B\,^1\Pi) = 17110.0(5)$ cm$^{-1}$. Using these values we obtain the value of $D_e(B\,^1\Pi)$ to be $1634.5(45)$ cm$^{-1}$, where the major source of error arises from the uncertainty in the value of $D_e(X\,^1\Sigma^+)$.

We also observed rotational perturbations in the $e$ parity levels of the $B\,^1\Pi$ ($v' = 2$) state, as manifested through irregularities in both P and R branches of the following $(v'', v')$ bands: $(v''=0, v'=2)$, $(v''=1, v'=2)$ and $(v''=2, v'=2)$. In addition, we observed that the energy splitting $\Delta_2 F_{v''}(J'')$ of the $X\,^1\Sigma^+$ state did not have any irregularities (Figure 4A). We did not observe any such irregularity in the $v' = 0$ level. These clearly indicate that the $B\,^1\Pi$ ($v' = 2$) level is perturbed, and that the $X\,^1\Sigma^+$ state is not. Figure 5 shows, for the $(v''=0, v'=2)$ band, the difference between observed and calculated line positions (calculated using $B_{v'}$ and $D_{v'}$ values obtained above from $f$ parity levels, i.e. Q branch, only). The maximum perturbation occurs at $J' \sim 16$. Since the $f$ parity levels (Q branch) are not perturbed, it can be concluded that the perturbing state is a $^1\Sigma^+$ state and not a $^1\Pi$ state [18]. From the *ab initio* calculations of potential energy curves [12], we expect that $C\,^1\Sigma^+$ state is the perturbing state. It is possible to extract the molecular constants of the $C\,^1\Sigma^+$ state from the nature and strength of the perturbation. However, our data set on perturbations is limited and the analysis is beyond the scope of this paper.

**Table 1.**
Molecular constants (in cm$^{-1}$) of the $X\,^1\Sigma^+$ state and $f$-parity levels of the $B\,^1\Pi$ state. All values of $T_v$ are referenced to $T_{v''=0} = 0$. The value of $D_{v''}$ is held fixed at $1.06 \times 10^{-6}$ cm$^{-1}$. These constants reproduce the observed transition frequencies with an accuracy of 0.02 cm$^{-1}$.

| $v$ | $T_v - T_{v''=0}$ | $B_v$ | $D_v \times 10^6$ |
|---|---|---|---|
| \multicolumn{4}{c}{$X\,^1\Sigma^+$ state} | | | |
| 0 | 0 | 0.21580 (3) | (fixed at |
| 1 | 192.96 | 0.21430 (5) | $1.06 \times 10^{-6}$) |
| 2 | 383.70 | 0.21265 (5) | |
| \multicolumn{4}{c}{$B\,^1\Pi$ state, $f$-levels} | | | |
| 0 | 17073.07 | 0.17146(4) | 1.6(1) |
| 1 | 17188.52 | 0.16703(4) | 1.2(1) |
| 2 | 17297.15 | 0.16261(2) | 1.78(5) |
| 3 | 17398.94 | 0.15768(4) | 2.00(8) |
| 4 | 17493.91 | 0.15226(6) | 1.7(2) |
| 5 | 17582.27 | 0.14717(2) | 2.17(5) |
| 6 | 17664.54 | 0.14194(3) | 2.42(6) |
| 7 | 17741.20 | 0.13696(8) | 3.4(4) |
| 8 | 17812.80 | 0.13135(4) | 2.07(7) |
| 9 | 17879.80 | 0.12660(8) | 2.2(2) |
| 10 | 17942.64 | 0.12191(5) | 2.2(1) |
| 11 | 18001.73 | 0.11743(3) | 2.29(6) |
| 12 | 18057.37 | 0.11301(6) | 2.2(2) |
| 13 | 18109.83 | 0.10907(2) | 2.61(5) |
| 14 | 18159.37 | 0.10488(4) | 2.35(9) |
| 15 | 18206.18 | 0.10113(6) | 2.7(1) |
| 16 | 18250.29 | 0.09738(10) | 1.7(4) |
| 17 | 18292.46 | 0.09364(13) | 1.9(5) |
| 18 | 18332.24 | 0.09026(1) | 2.09(3) |
| 19 | 18369.95 | 0.08696(3) | 2.33(4) |
| 20 | 18405.63 | 0.08349(3) | 2.31(6) |

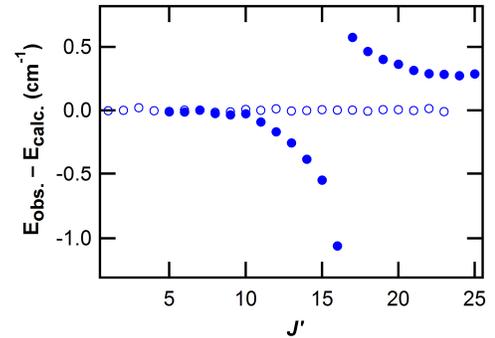

**Figure 5.** The difference between observed and calculated transition frequencies in the $(v'' = 0, v' = 2)$ band of the $X\,^1\Sigma^+ \to B\,^1\Pi$ system. Open circles (○) correspond to $f$ parity levels and filled circles (●) correspond to $e$ parity levels. The $f$ parity levels are not perturbed while rotational perturbations are maximum near $J' = 16$ of the $e$ parity levels of the $B\,^1\Pi$ state.

Before we conclude, we look back at the relative intensity of the LIF lines in Figure 2C. The line originating from $v' = 2$ or $v' = 3$ and terminating in $v'' = 0$ has much higher intensity than those terminating in $v'' > 0$. This implies that the Franck Condon factor is very high for the $X\,^1\Sigma^+$ ($v'' = 0$) ← $B\,^1\Pi$ ($v' = 2$) and $X\,^1\Sigma^+$ ($v'' = 0$) ← $B\,^1\Pi$ ($v' = 3$) transitions. A rough estimate based on the relative intensity of the LIF peaks shows that approximately 25% of the molecules in the $v' = 2$ decay back to $v'' = 0$ and approximately 30% of the molecules in $v' = 3$ states decay back to $v'' = 0$. This is encouraging for formation of ultra-cold ground state molecules via photoassociation [14,15] into the $B\,^1\Pi$ ($v' = 2$) or $B\,^1\Pi$ ($v' = 3$) state followed by spontaneous emission to the ground $X\,^1\Sigma^+$ ($v'' = 0$) state. A similar scheme was used for producing ultra-cold LiCs molecules [15]. In particular, $B\,^1\Pi$ ($v' = 3$) could be an excellent state for such a scheme, especially if there is significant increase in the photoassociation efficiency due to wave function overlap between the $X\,^1\Sigma^+$ and $a\,^3\Sigma^+$ states near the inner turning point of the $a\,^3\Sigma^+$ state. In the absence of such overlap, the recently proposed scheme based on photoassociation near a Feshbach resonance [19] could be used to form ultracold LiRb molecules in the ground $X\,^1\Sigma^+$ ($v'' = 0$) state. It is also clear that the $B\,^1\Pi$ ($v' = 2$) or $B\,^1\Pi$ ($v' = 3$) state would be a good intermediate state for state selective detection of ultra-cold ground state LiRb molecules through Resonance Enhanced Multi Photon Ionization (REMPI).

## 4. Conclusion

In conclusion, we have determined the molecular constants for the $B\,^1\Pi$ state of $^7$Li$^{85}$Rb for the first time. We find slight disagreement between our experimentally determined constants for the $B\,^1\Pi$ state and those reported by Korek *et al.* [12] based on *ab initio* calculations. For the $B\,^1\Pi$ state, ref. [12] overestimates the value of $T_e$ by only ~ 95 cm$^{-1}$, $\omega_e$ by ~ 17 cm$^{-1}$, and $B_e$ by ~ 0.006 cm$^{-1}$. We have also determined the molecular constants for the first three vibrational levels of the $X\,^1\Sigma^+$ state and find good agreement with the recently reported experimental values [10]. We observed rotational perturbations in the $v' = 2$ level of the $B\,^1\Pi$ state and identified the perturbing state. We find that $v' = 2$ and $v' = 3$ levels of the $B\,^1\Pi$ state would be good intermediate state for state selective detection of ultra-cold ground state LiRb molecules. We also predict that $v' = 3$ level of the $B\,^1\Pi$ state would be an excellent intermediate state for formation of LiRb molecules in the ground $v'' = 0$, $J'' = 0$ level of the $X\,^1\Sigma^+$ state by photoassociation.

## Acknowledgment

This work was partially supported by the National Science Foundation (grant CCF0829918) and the Army Research Office (grant W911NF-10-1-0243). We gratefully acknowledge A. M. Weiner, for the loan of the monochromator; and M. Korek and A. R. Allouche, for providing the theoretical potential energy curves.

## Supplementary material

Supplementary data associated with this article can be found after the references.

# Supplementary Material

# Laser spectroscopy of the $X\,^1\Sigma^+$ and $B\,^1\Pi$ states of LiRb molecule

**Sourav Dutta[1,*], Adeel Altaf[1], D. S. Elliott[1,2] and Yong P. Chen[1,2]**

[1]*Department of Physics, Purdue University, West Lafayette, IN 47907, USA*
[2]*School of Electrical and Computer Engineering, Purdue University, West Lafayette, IN 47907, USA*
**\*** *sourav.dutta.mr@gmail.com*

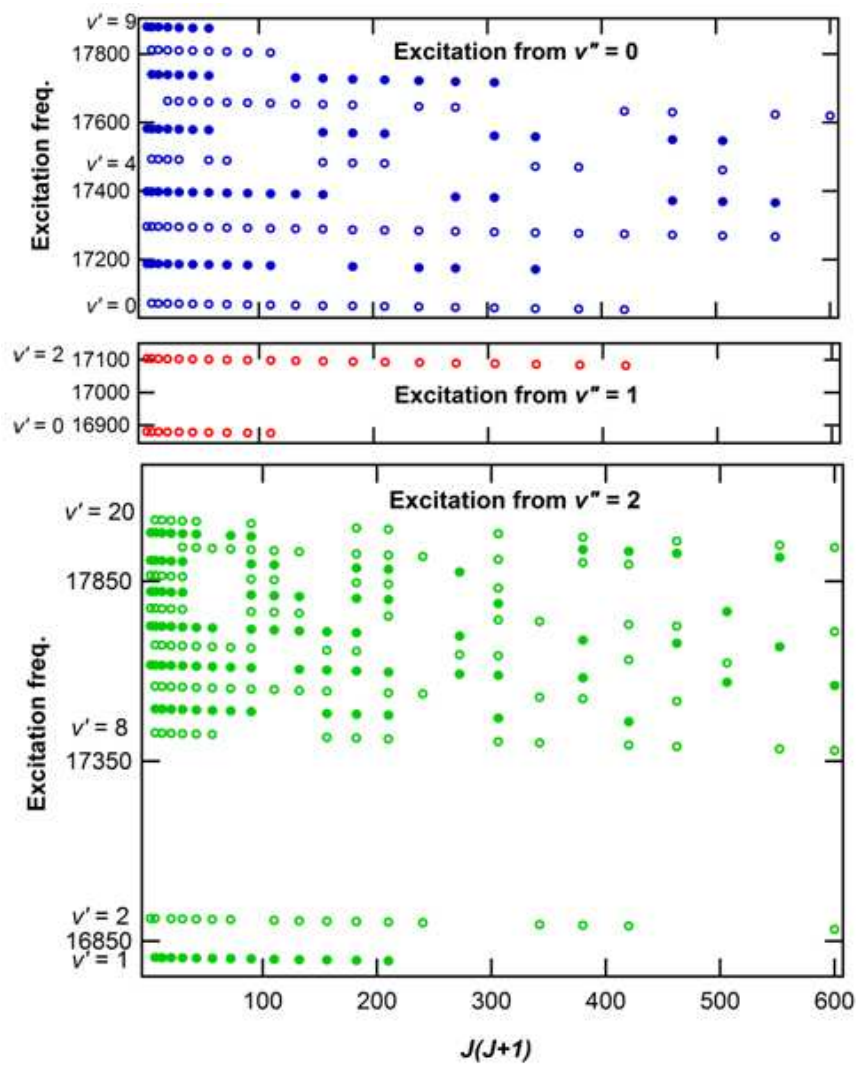

**Figure S1.** The transition frequencies of all $Q$ branch transitions observed in this study plotted against $J(J+1)$. For clarity, even and odd $v'$ levels are denoted by open circles (○) and filled circles (●) respectively.

**Table S1.**

List of all experimentally observed transitions in $^7$Li$^{85}$Rb using Fluorescence Excitation Spectroscopy (FES). The average accuracy of the line positions is 0.02 cm$^{-1}$.

Notations:

$v''$ and $v'$ are the vibrational quantum numbers of the $X\ ^1\Sigma^+$ and $B\ ^1\Pi$ states respectively.

$J''$ and $J'$ are the rotational quantum numbers of the $X\ ^1\Sigma^+$ and $B\ ^1\Pi$ states respectively.

| Number | $v''$ | $J''$ | $v'$ | $J'$ | Frequency (cm$^{-1}$) |
|---|---|---|---|---|---|
| 1 | 0 | 2 | 0 | 2 | 17072.63 |
| 2 | 0 | 3 | 0 | 3 | 17072.36 |
| 3 | 0 | 4 | 0 | 4 | 17072.01 |
| 4 | 0 | 5 | 0 | 5 | 17071.58 |
| 5 | 0 | 6 | 0 | 6 | 17071.05 |
| 6 | 0 | 7 | 0 | 7 | 17070.41 |
| 7 | 0 | 8 | 0 | 8 | 17069.71 |
| 8 | 0 | 9 | 0 | 9 | 17068.90 |
| 9 | 0 | 10 | 0 | 10 | 17067.99 |
| 10 | 0 | 11 | 0 | 11 | 17067.03 |
| 11 | 0 | 12 | 0 | 12 | 17065.98 |
| 12 | 0 | 13 | 0 | 13 | 17064.82 |
| 13 | 0 | 14 | 0 | 14 | 17063.57 |
| 14 | 0 | 15 | 0 | 15 | 17062.23 |
| 15 | 0 | 16 | 0 | 16 | 17060.81 |
| 16 | 0 | 17 | 0 | 17 | 17059.29 |
| 17 | 0 | 18 | 0 | 18 | 17057.69 |
| 18 | 0 | 19 | 0 | 19 | 17056.00 |
| 19 | 0 | 20 | 0 | 20 | 17054.17 |
| 20 | 0 | 1 | 1 | 1 | 17188.24 |
| 21 | 0 | 2 | 1 | 2 | 17188.07 |
| 22 | 0 | 3 | 1 | 3 | 17187.77 |
| 23 | 0 | 4 | 1 | 4 | 17187.39 |
| 24 | 0 | 5 | 1 | 5 | 17186.89 |
| 25 | 0 | 6 | 1 | 6 | 17186.29 |
| 26 | 0 | 7 | 1 | 7 | 17185.62 |
| 27 | 0 | 8 | 1 | 8 | 17184.85 |
| 28 | 0 | 9 | 1 | 9 | 17183.96 |
| 29 | 0 | 10 | 1 | 10 | 17182.99 |
| 30 | 0 | 13 | 1 | 13 | 17179.47 |
| 31 | 0 | 15 | 1 | 15 | 17176.64 |
| 32 | 0 | 16 | 1 | 16 | 17175.08 |
| 33 | 0 | 18 | 1 | 18 | 17171.66 |
| 34 | 0 | 1 | 2 | 1 | 17296.88 |
| 35 | 0 | 2 | 2 | 2 | 17296.67 |
| 36 | 0 | 3 | 2 | 3 | 17296.37 |
| 37 | 0 | 4 | 2 | 4 | 17295.92 |
| 38 | 0 | 5 | 2 | 5 | 17295.38 |
| 39 | 0 | 6 | 2 | 6 | 17294.76 |
| 40 | 0 | 7 | 2 | 7 | 17294.01 |
| 41 | 0 | 8 | 2 | 8 | 17293.14 |
| 42 | 0 | 9 | 2 | 9 | 17292.18 |
| 43 | 0 | 10 | 2 | 10 | 17291.14 |
| 44 | 0 | 11 | 2 | 11 | 17289.96 |
| 45 | 0 | 12 | 2 | 12 | 17288.69 |
| 46 | 0 | 13 | 2 | 13 | 17287.28 |
| 47 | 0 | 14 | 2 | 14 | 17285.79 |
| 48 | 0 | 15 | 2 | 15 | 17284.19 |
| 49 | 0 | 16 | 2 | 16 | 17282.47 |
| 50 | 0 | 17 | 2 | 17 | 17280.65 |
| 51 | 0 | 18 | 2 | 18 | 17278.71 |
| 52 | 0 | 19 | 2 | 19 | 17276.68 |
| 53 | 0 | 20 | 2 | 20 | 17274.53 |
| 54 | 0 | 21 | 2 | 21 | 17272.26 |
| 55 | 0 | 22 | 2 | 22 | 17269.90 |
| 56 | 0 | 23 | 2 | 23 | 17267.40 |
| 57 | 0 | 1 | 3 | 1 | 17398.66 |
| 58 | 0 | 2 | 3 | 2 | 17398.45 |
| 59 | 0 | 3 | 3 | 3 | 17398.08 |
| 60 | 0 | 4 | 3 | 4 | 17397.63 |
| 61 | 0 | 5 | 3 | 5 | 17397.05 |
| 62 | 0 | 6 | 3 | 6 | 17396.36 |
| 63 | 0 | 7 | 3 | 7 | 17395.54 |
| 64 | 0 | 8 | 3 | 8 | 17394.60 |
| 65 | 0 | 9 | 3 | 9 | 17393.54 |
| 66 | 0 | 10 | 3 | 10 | 17392.39 |
| 67 | 0 | 11 | 3 | 11 | 17391.09 |
| 68 | 0 | 12 | 3 | 12 | 17389.67 |
| 69 | 0 | 16 | 3 | 16 | 17382.90 |
| 70 | 0 | 17 | 3 | 17 | 17380.91 |
| 71 | 0 | 21 | 3 | 21 | 17371.73 |
| 72 | 0 | 22 | 3 | 22 | 17369.13 |
| 73 | 0 | 23 | 3 | 23 | 17366.42 |
| 74 | 0 | 2 | 4 | 2 | 17493.37 |
| 75 | 0 | 3 | 4 | 3 | 17492.98 |
| 76 | 0 | 4 | 4 | 4 | 17492.49 |
| 77 | 0 | 5 | 4 | 5 | 17491.83 |
| 78 | 0 | 7 | 4 | 7 | 17490.19 |
| 79 | 0 | 8 | 4 | 8 | 17489.15 |
| 80 | 0 | 12 | 4 | 12 | 17483.83 |
| 81 | 0 | 13 | 4 | 13 | 17482.18 |
| 82 | 0 | 14 | 4 | 14 | 17480.38 |
| 83 | 0 | 18 | 4 | 18 | 17471.95 |
| 84 | 0 | 19 | 4 | 19 | 17469.54 |
| 85 | 0 | 22 | 4 | 22 | 17461.42 |
| 86 | 0 | 1 | 5 | 1 | 17581.99 |
| 87 | 0 | 2 | 5 | 2 | 17581.71 |
| 88 | 0 | 3 | 5 | 3 | 17581.31 |
| 89 | 0 | 4 | 5 | 4 | 17580.75 |
| 90 | 0 | 5 | 5 | 5 | 17580.07 |
| 91 | 0 | 6 | 5 | 6 | 17579.23 |
| 92 | 0 | 7 | 5 | 7 | 17578.28 |
| 93 | 0 | 12 | 5 | 12 | 17571.39 |
| 94 | 0 | 13 | 5 | 13 | 17569.60 |
| 95 | 0 | 14 | 5 | 14 | 17567.67 |
| 96 | 0 | 17 | 5 | 17 | 17561.02 |
| 97 | 0 | 18 | 5 | 18 | 17558.52 |
| 98 | 0 | 21 | 5 | 21 | 17550.17 |
| 99 | 0 | 22 | 5 | 22 | 17547.12 |
| 100 | 0 | 4 | 6 | 4 | 17662.91 |
| 101 | 0 | 5 | 6 | 5 | 17662.20 |
| 102 | 0 | 6 | 6 | 6 | 17661.29 |
| 103 | 0 | 7 | 6 | 7 | 17660.26 |
| 104 | 0 | 8 | 6 | 8 | 17659.08 |
| 105 | 0 | 9 | 6 | 9 | 17657.74 |
| 106 | 0 | 10 | 6 | 10 | 17656.24 |
| 107 | 0 | 11 | 6 | 11 | 17654.62 |
| 108 | 0 | 12 | 6 | 12 | 17652.84 |
| 109 | 0 | 13 | 6 | 13 | 17650.91 |
| 110 | 0 | 15 | 6 | 15 | 17646.61 |
| 111 | 0 | 16 | 6 | 16 | 17644.22 |
| 112 | 0 | 20 | 6 | 20 | 17633.14 |
| 113 | 0 | 21 | 6 | 21 | 17629.97 |
| 114 | 0 | 23 | 6 | 23 | 17623.23 |
| 115 | 0 | 24 | 6 | 24 | 17619.59 |
| 116 | 0 | 2 | 7 | 2 | 17740.59 |
| 117 | 0 | 3 | 7 | 3 | 17740.12 |
| 118 | 0 | 4 | 7 | 4 | 17739.49 |
| 119 | 0 | 5 | 7 | 5 | 17738.71 |
| 120 | 0 | 6 | 7 | 6 | 17737.76 |
| 121 | 0 | 7 | 7 | 7 | 17736.63 |
| 122 | 0 | 11 | 7 | 11 | 17730.62 |
| 123 | 0 | 12 | 7 | 12 | 17728.71 |
| 124 | 0 | 13 | 7 | 13 | 17726.63 |
| 125 | 0 | 14 | 7 | 14 | 17724.40 |
| 126 | 0 | 15 | 7 | 15 | 17722.01 |
| 127 | 0 | 16 | 7 | 16 | 17719.44 |
| 128 | 0 | 17 | 7 | 17 | 17716.74 |
| 129 | 0 | 2 | 8 | 2 | 17812.17 |
| 130 | 0 | 3 | 8 | 3 | 17811.66 |
| 131 | 0 | 4 | 8 | 4 | 17810.96 |
| 132 | 0 | 5 | 8 | 5 | 17810.14 |
| 133 | 0 | 6 | 8 | 6 | 17809.12 |
| 134 | 0 | 7 | 8 | 7 | 17807.95 |
| 135 | 0 | 8 | 8 | 8 | 17806.62 |
| 136 | 0 | 9 | 8 | 9 | 17805.10 |
| 137 | 0 | 10 | 8 | 10 | 17803.38 |
| 138 | 0 | 1 | 9 | 1 | 17879.49 |
| 139 | 0 | 2 | 9 | 2 | 17879.14 |
| 140 | 0 | 3 | 9 | 3 | 17878.60 |
| 141 | 0 | 4 | 9 | 4 | 17877.89 |
| 142 | 0 | 5 | 9 | 5 | 17877.00 |
| 143 | 0 | 6 | 9 | 6 | 17875.91 |
| 144 | 0 | 7 | 9 | 7 | 17874.70 |
| 145 | 1 | 1 | 0 | 1 | 16879.86 |
| 146 | 1 | 2 | 0 | 2 | 16879.69 |
| 147 | 1 | 3 | 0 | 3 | 16879.43 |
| 148 | 1 | 4 | 0 | 4 | 16879.09 |
| 149 | 1 | 5 | 0 | 5 | 16878.66 |
| 150 | 1 | 6 | 0 | 6 | 16878.15 |
| 151 | 1 | 7 | 0 | 7 | 16877.52 |
| 152 | 1 | 8 | 0 | 8 | 16876.84 |
| 153 | 1 | 9 | 0 | 9 | 16876.07 |
| 154 | 1 | 10 | 0 | 10 | 16875.22 |
| 155 | 1 | 1 | 2 | 1 | 17103.93 |
| 156 | 1 | 2 | 2 | 2 | 17103.71 |
| 157 | 1 | 3 | 2 | 3 | 17103.43 |
| 158 | 1 | 4 | 2 | 4 | 17102.99 |
| 159 | 1 | 5 | 2 | 5 | 17102.47 |
| 160 | 1 | 6 | 2 | 6 | 17101.85 |
| 161 | 1 | 7 | 2 | 7 | 17101.15 |
| 162 | 1 | 8 | 2 | 8 | 17100.30 |
| 163 | 1 | 9 | 2 | 9 | 17099.40 |
| 164 | 1 | 10 | 2 | 10 | 17098.37 |
| 165 | 1 | 11 | 2 | 11 | 17097.21 |
| 166 | 1 | 12 | 2 | 12 | 17095.96 |
| 167 | 1 | 13 | 2 | 13 | 17094.61 |
| 168 | 1 | 14 | 2 | 14 | 17093.16 |
| 169 | 1 | 15 | 2 | 15 | 17091.61 |
| 170 | 1 | 16 | 2 | 16 | 17089.93 |
| 171 | 1 | 17 | 2 | 17 | 17088.16 |
| 172 | 1 | 18 | 2 | 18 | 17086.28 |
| 173 | 1 | 19 | 2 | 19 | 17084.30 |
| 174 | 1 | 20 | 2 | 20 | 17082.21 |
| 175 | 2 | 2 | 1 | 2 | 16804.35 |
| 176 | 2 | 3 | 1 | 3 | 16804.10 |
| 177 | 2 | 4 | 1 | 4 | 16803.73 |
| 178 | 2 | 5 | 1 | 5 | 16803.28 |
| 179 | 2 | 6 | 1 | 6 | 16802.71 |
| 180 | 2 | 7 | 1 | 7 | 16802.09 |
| 181 | 2 | 8 | 1 | 8 | 16801.36 |
| 182 | 2 | 9 | 1 | 9 | 16800.54 |
| 183 | 2 | 10 | 1 | 10 | 16799.61 |
| 184 | 2 | 11 | 1 | 11 | 16798.62 |
| 185 | 2 | 12 | 1 | 12 | 16797.52 |
| 186 | 2 | 13 | 1 | 13 | 16796.31 |
| 187 | 2 | 14 | 1 | 14 | 16795.07 |
| 188 | 2 | 1 | 2 | 1 | 16913.22 |
| 189 | 2 | 2 | 2 | 2 | 16913.02 |
| 190 | 2 | 4 | 2 | 4 | 16912.32 |
| 191 | 2 | 5 | 2 | 5 | 16911.82 |
| 192 | 2 | 6 | 2 | 6 | 16911.21 |
| 193 | 2 | 7 | 2 | 7 | 16910.51 |
| 194 | 2 | 8 | 2 | 8 | 16909.70 |
| 195 | 2 | 10 | 2 | 10 | 16907.79 |
| 196 | 2 | 11 | 2 | 11 | 16906.69 |
| 197 | 2 | 12 | 2 | 12 | 16905.48 |
| 198 | 2 | 13 | 2 | 13 | 16904.16 |
| 199 | 2 | 14 | 2 | 14 | 16902.76 |
| 200 | 2 | 15 | 2 | 15 | 16901.24 |
| 201 | 2 | 18 | 2 | 18 | 16896.09 |
| 202 | 2 | 19 | 2 | 19 | 16894.16 |
| 203 | 2 | 20 | 2 | 20 | 16892.13 |
| 204 | 2 | 2 | 8 | 2 | 17428.50 |
| 205 | 2 | 3 | 8 | 3 | 17428.01 |
| 206 | 2 | 4 | 8 | 4 | 17427.34 |
| 207 | 2 | 5 | 8 | 5 | 17426.52 |
| 208 | 2 | 6 | 8 | 6 | 17425.55 |
| 209 | 2 | 7 | 8 | 7 | 17424.43 |
| 210 | 2 | 12 | 8 | 12 | 17416.26 |
| 211 | 2 | 13 | 8 | 13 | 17414.14 |
| 212 | 2 | 14 | 8 | 14 | 17411.87 |
| 213 | 2 | 17 | 8 | 17 | 17404.02 |
| 214 | 2 | 18 | 8 | 18 | 17401.05 |
| 215 | 2 | 20 | 8 | 20 | 17394.66 |
| 216 | 2 | 21 | 8 | 21 | 17391.21 |
| 217 | 2 | 23 | 8 | 23 | 17383.81 |
| 218 | 2 | 24 | 8 | 24 | 17379.82 |
| 219 | 2 | 2 | 9 | 2 | 17495.46 |
| 220 | 2 | 3 | 9 | 3 | 17494.91 |
| 221 | 2 | 4 | 9 | 4 | 17494.23 |
| 222 | 2 | 5 | 9 | 5 | 17493.37 |
| 223 | 2 | 6 | 9 | 6 | 17492.37 |
| 224 | 2 | 7 | 9 | 7 | 17491.14 |
| 225 | 2 | 8 | 9 | 8 | 17489.76 |
| 226 | 2 | 9 | 9 | 9 | 17488.22 |
| 227 | 2 | 12 | 9 | 12 | 17482.49 |
| 228 | 2 | 13 | 9 | 13 | 17480.29 |
| 229 | 2 | 14 | 9 | 14 | 17477.84 |
| 230 | 2 | 17 | 9 | 17 | 17469.54 |
| 231 | 2 | 20 | 9 | 20 | 17459.63 |
| 232 | 2 | 2 | 10 | 2 | 17558.28 |
| 233 | 2 | 3 | 10 | 3 | 17557.74 |
| 234 | 2 | 4 | 10 | 4 | 17557.01 |
| 235 | 2 | 5 | 10 | 5 | 17556.09 |
| 236 | 2 | 6 | 10 | 6 | 17555.01 |
| 237 | 2 | 7 | 10 | 7 | 17553.71 |
| 238 | 2 | 8 | 10 | 8 | 17552.30 |
| 239 | 2 | 9 | 10 | 9 | 17550.63 |
| 240 | 2 | 10 | 10 | 10 | 17548.82 |
| 241 | 2 | 11 | 10 | 11 | 17546.82 |
| 242 | 2 | 12 | 10 | 12 | 17544.63 |
| 243 | 2 | 14 | 10 | 14 | 17539.74 |
| 244 | 2 | 15 | 10 | 15 | 17536.97 |
| 245 | 2 | 18 | 10 | 18 | 17527.65 |
| 246 | 2 | 19 | 10 | 19 | 17524.18 |
| 247 | 2 | 21 | 10 | 21 | 17516.66 |
| 248 | 2 | 1 | 11 | 1 | 17617.73 |
| 249 | 2 | 2 | 11 | 2 | 17617.33 |
| 250 | 2 | 3 | 11 | 3 | 17616.77 |
| 251 | 2 | 4 | 11 | 4 | 17616.02 |
| 252 | 2 | 5 | 11 | 5 | 17615.06 |
| 253 | 2 | 6 | 11 | 6 | 17613.91 |
| 254 | 2 | 7 | 11 | 7 | 17612.58 |
| 255 | 2 | 8 | 11 | 8 | 17611.03 |
| 256 | 2 | 9 | 11 | 9 | 17609.32 |
| 257 | 2 | 11 | 11 | 11 | 17605.32 |
| 258 | 2 | 12 | 11 | 12 | 17603.03 |
| 259 | 2 | 13 | 11 | 13 | 17600.54 |
| 260 | 2 | 14 | 11 | 14 | 17597.86 |
| 261 | 2 | 16 | 11 | 16 | 17591.94 |
| 262 | 2 | 17 | 11 | 17 | 17588.66 |
| 263 | 2 | 19 | 11 | 19 | 17581.55 |
| 264 | 2 | 22 | 11 | 22 | 17569.41 |
| 265 | 2 | 24 | 11 | 24 | 17560.34 |
| 266 | 2 | 2 | 12 | 2 | 17672.97 |
| 267 | 2 | 3 | 12 | 3 | 17672.37 |
| 268 | 2 | 4 | 12 | 4 | 17671.56 |
| 269 | 2 | 5 | 12 | 5 | 17670.56 |
| 270 | 2 | 6 | 12 | 6 | 17669.37 |
| 271 | 2 | 7 | 12 | 7 | 17667.97 |
| 272 | 2 | 8 | 12 | 8 | 17666.38 |
| 273 | 2 | 9 | 12 | 9 | 17664.54 |
| 274 | 2 | 12 | 12 | 12 | 17657.99 |
| 275 | 2 | 13 | 12 | 13 | 17655.40 |
| 276 | 2 | 16 | 12 | 16 | 17646.39 |
| 277 | 2 | 17 | 12 | 17 | 17642.97 |
| 278 | 2 | 20 | 12 | 20 | 17631.52 |
| 279 | 2 | 22 | 12 | 22 | 17622.85 |
| 280 | 2 | 1 | 13 | 1 | 17725.81 |
| 281 | 2 | 2 | 13 | 2 | 17725.41 |
| 282 | 2 | 3 | 13 | 3 | 17724.78 |
| 283 | 2 | 4 | 13 | 4 | 17723.96 |
| 284 | 2 | 5 | 13 | 5 | 17722.92 |
| 285 | 2 | 6 | 13 | 6 | 17721.67 |
| 286 | 2 | 7 | 13 | 7 | 17720.22 |
| 287 | 2 | 9 | 13 | 9 | 17716.69 |
| 288 | 2 | 10 | 13 | 10 | 17714.60 |
| 289 | 2 | 11 | 13 | 11 | 17712.31 |
| 290 | 2 | 12 | 13 | 12 | 17709.83 |
| 291 | 2 | 13 | 13 | 13 | 17707.11 |
| 292 | 2 | 16 | 13 | 16 | 17697.74 |
| 293 | 2 | 19 | 13 | 19 | 17686.44 |
| 294 | 2 | 21 | 13 | 21 | 17677.84 |
| 295 | 2 | 23 | 13 | 23 | 17668.38 |
| 296 | 2 | 1 | 14 | 1 | 17775.34 |
| 297 | 2 | 2 | 14 | 2 | 17774.93 |
| 298 | 2 | 3 | 14 | 3 | 17774.27 |
| 299 | 2 | 4 | 14 | 4 | 17773.42 |
| 300 | 2 | 5 | 14 | 5 | 17772.31 |
| 301 | 2 | 9 | 14 | 9 | 17765.84 |
| 302 | 2 | 10 | 14 | 10 | 17763.69 |
| 303 | 2 | 11 | 14 | 11 | 17761.31 |
| 304 | 2 | 14 | 14 | 14 | 17752.88 |
| 305 | 2 | 17 | 14 | 17 | 17742.48 |
| 306 | 2 | 18 | 14 | 18 | 17738.55 |
| 307 | 2 | 20 | 14 | 20 | 17730.07 |
| 308 | 2 | 21 | 14 | 21 | 17725.53 |
| 309 | 2 | 24 | 14 | 24 | 17710.43 |
| 310 | 2 | 1 | 15 | 1 | 17822.14 |
| 311 | 2 | 2 | 15 | 2 | 17821.72 |
| 312 | 2 | 3 | 15 | 3 | 17821.06 |
| 313 | 2 | 4 | 15 | 4 | 17820.17 |
| 314 | 2 | 5 | 15 | 5 | 17819.05 |
| 315 | 2 | 9 | 15 | 9 | 17812.33 |
| 316 | 2 | 10 | 15 | 10 | 17810.07 |
| 317 | 2 | 11 | 15 | 11 | 17807.63 |
| 318 | 2 | 13 | 15 | 13 | 17802.02 |
| 319 | 2 | 14 | 15 | 14 | 17798.88 |
| 320 | 2 | 17 | 15 | 17 | 17788.12 |
| 321 | 2 | 22 | 15 | 22 | 17765.52 |
| 322 | 2 | 1 | 16 | 1 | 17866.27 |
| 323 | 2 | 2 | 16 | 2 | 17865.79 |
| 324 | 2 | 3 | 16 | 3 | 17865.09 |
| 325 | 2 | 4 | 16 | 4 | 17864.19 |
| 326 | 2 | 5 | 16 | 5 | 17863.04 |
| 327 | 2 | 9 | 16 | 9 | 17856.09 |
| 328 | 2 | 10 | 16 | 10 | 17853.80 |
| 329 | 2 | 13 | 16 | 13 | 17845.51 |
| 330 | 2 | 14 | 16 | 14 | 17842.26 |
| 331 | 2 | 17 | 16 | 17 | 17831.16 |
| 332 | 2 | 1 | 17 | 1 | 17908.42 |
| 333 | 2 | 2 | 17 | 2 | 17907.96 |
| 334 | 2 | 3 | 17 | 3 | 17907.23 |
| 335 | 2 | 4 | 17 | 4 | 17906.28 |
| 336 | 2 | 5 | 17 | 5 | 17905.09 |
| 337 | 2 | 9 | 17 | 9 | 17897.95 |
| 338 | 2 | 10 | 17 | 10 | 17895.56 |
| 339 | 2 | 13 | 17 | 13 | 17886.98 |
| 340 | 2 | 14 | 17 | 14 | 17883.64 |
| 341 | 2 | 16 | 17 | 16 | 17876.23 |
| 342 | 2 | 5 | 18 | 5 | 17944.78 |
| 343 | 2 | 6 | 18 | 6 | 17943.31 |
| 344 | 2 | 7 | 18 | 7 | 17941.59 |
| 345 | 2 | 8 | 18 | 8 | 17939.63 |
| 346 | 2 | 9 | 18 | 9 | 17937.42 |
| 347 | 2 | 10 | 18 | 10 | 17934.97 |
| 348 | 2 | 11 | 18 | 11 | 17932.28 |
| 349 | 2 | 13 | 18 | 13 | 17926.14 |
| 350 | 2 | 14 | 18 | 14 | 17922.70 |
| 351 | 2 | 15 | 18 | 15 | 17919.02 |
| 352 | 2 | 17 | 18 | 17 | 17910.90 |
| 353 | 2 | 19 | 18 | 19 | 17901.80 |
| 354 | 2 | 20 | 18 | 20 | 17896.86 |
| 355 | 2 | 1 | 19 | 1 | 17985.92 |
| 356 | 2 | 2 | 19 | 2 | 17985.40 |
| 357 | 2 | 3 | 19 | 3 | 17984.65 |
| 358 | 2 | 4 | 19 | 4 | 17983.64 |
| 359 | 2 | 5 | 19 | 5 | 17982.41 |
| 360 | 2 | 6 | 19 | 6 | 17980.89 |
| 361 | 2 | 8 | 19 | 8 | 17977.10 |
| 362 | 2 | 9 | 19 | 9 | 17974.83 |
| 363 | 2 | 19 | 19 | 19 | 17938.22 |
| 364 | 2 | 20 | 19 | 20 | 17933.16 |
| 365 | 2 | 21 | 19 | 21 | 17927.73 |
| 366 | 2 | 23 | 19 | 23 | 17916.39 |
| 367 | 2 | 25 | 19 | 25 | 17903.92 |
| 368 | 2 | 27 | 19 | 27 | 17890.43 |
| 369 | 2 | 28 | 19 | 28 | 17883.27 |
| 370 | 2 | 2 | 20 | 2 | 18021.08 |
| 371 | 2 | 3 | 20 | 3 | 18020.30 |
| 372 | 2 | 4 | 20 | 4 | 18019.26 |
| 373 | 2 | 5 | 20 | 5 | 18017.96 |
| 374 | 2 | 6 | 20 | 6 | 18016.42 |
| 375 | 2 | 9 | 20 | 9 | 18010.21 |
| 376 | 2 | 13 | 20 | 13 | 17998.30 |
| 377 | 2 | 14 | 20 | 14 | 17994.67 |
| 378 | 2 | 17 | 20 | 17 | 17982.22 |
| 379 | 2 | 19 | 20 | 19 | 17972.58 |
| 380 | 2 | 21 | 20 | 21 | 17961.90 |
| 381 | 2 | 23 | 20 | 23 | 17950.16 |
| 382 | 2 | 24 | 20 | 24 | 17943.91 |
| 383 | 0 | 3 | 0 | 4 | 17073.73 |
| 384 | 0 | 4 | 0 | 5 | 17073.73 |
| 385 | 0 | 5 | 0 | 6 | 17073.62 |
| 386 | 0 | 6 | 0 | 7 | 17073.41 |
| 387 | 0 | 7 | 0 | 8 | 17073.15 |
| 388 | 0 | 8 | 0 | 9 | 17072.80 |
| 389 | 0 | 9 | 0 | 10 | 17072.36 |
| 390 | 0 | 10 | 0 | 11 | 17071.78 |
| 391 | 0 | 11 | 0 | 12 | 17071.14 |
| 392 | 0 | 12 | 0 | 13 | 17070.41 |
| 393 | 0 | 13 | 0 | 14 | 17069.59 |
| 394 | 0 | 14 | 0 | 15 | 17068.66 |
| 395 | 0 | 15 | 0 | 16 | 17067.67 |
| 396 | 0 | 16 | 0 | 17 | 17066.59 |
| 397 | 0 | 17 | 0 | 18 | 17065.40 |
| 398 | 0 | 18 | 0 | 19 | 17064.12 |
| 399 | 0 | 19 | 0 | 20 | 17062.78 |
| 400 | 0 | 20 | 0 | 21 | 17061.32 |
| 401 | 0 | 21 | 0 | 22 | 17059.78 |
| 402 | 0 | 22 | 0 | 23 | 17058.15 |
| 403 | 0 | 23 | 0 | 24 | 17056.41 |
| 404 | 0 | 24 | 0 | 25 | 17054.57 |
| 405 | 0 | 25 | 0 | 26 | 17052.65 |
| 406 | 0 | 5 | 0 | 4 | 17069.86 |
| 407 | 0 | 6 | 0 | 5 | 17068.98 |
| 408 | 0 | 7 | 0 | 6 | 17068.02 |
| 409 | 0 | 8 | 0 | 7 | 17066.96 |
| 410 | 0 | 9 | 0 | 8 | 17065.81 |
| 411 | 0 | 10 | 0 | 9 | 17064.58 |
| 412 | 0 | 11 | 0 | 10 | 17063.27 |
| 413 | 0 | 12 | 0 | 11 | 17061.85 |
| 414 | 0 | 13 | 0 | 12 | 17060.36 |
| 415 | 0 | 14 | 0 | 13 | 17058.76 |
| 416 | 0 | 15 | 0 | 14 | 17057.10 |
| 417 | 0 | 16 | 0 | 15 | 17055.33 |
| 418 | 0 | 17 | 0 | 16 | 17053.47 |
| 419 | 0 | 18 | 0 | 17 | 17051.55 |
| 420 | 0 | 19 | 0 | 18 | 17049.48 |
| 421 | 0 | 23 | 0 | 22 | 17040.45 |
| 422 | 0 | 27 | 0 | 26 | 17029.94 |
| 423 | 1 | 5 | 0 | 6 | 16880.69 |
| 424 | 1 | 6 | 0 | 7 | 16880.53 |
| 425 | 1 | 7 | 0 | 8 | 16880.26 |
| 426 | 1 | 8 | 0 | 9 | 16879.92 |
| 427 | 1 | 9 | 0 | 10 | 16879.49 |
| 428 | 1 | 10 | 0 | 11 | 16878.98 |
| 429 | 1 | 11 | 0 | 12 | 16878.36 |
| 430 | 1 | 12 | 0 | 13 | 16877.68 |
| 431 | 1 | 14 | 0 | 15 | 16876.04 |
| 432 | 1 | 15 | 0 | 16 | 16875.07 |
| 433 | 1 | 16 | 0 | 17 | 16874.02 |
| 434 | 1 | 17 | 0 | 18 | 16872.88 |
| 435 | 1 | 7 | 0 | 6 | 16875.13 |
| 436 | 1 | 8 | 0 | 7 | 16874.11 |
| 437 | 1 | 9 | 0 | 8 | 16872.97 |
| 438 | 1 | 10 | 0 | 9 | 16871.80 |
| 439 | 1 | 11 | 0 | 10 | 16870.49 |
| 440 | 1 | 12 | 0 | 11 | 16869.12 |
| 441 | 1 | 13 | 0 | 12 | 16867.66 |
| 442 | 1 | 16 | 0 | 15 | 16862.77 |
| 443 | 1 | 17 | 0 | 16 | 16860.96 |
| 444 | 1 | 18 | 0 | 17 | 16859.06 |
| 445 | 1 | 19 | 0 | 18 | 16857.10 |
| 446 | 2 | 13 | 0 | 14 | 16686.44 |
| 447 | 2 | 14 | 0 | 15 | 16685.63 |
| 448 | 2 | 15 | 0 | 16 | 16684.71 |
| 449 | 2 | 16 | 0 | 17 | 16683.71 |
| 450 | 2 | 17 | 0 | 18 | 16682.63 |
| 451 | 2 | 18 | 0 | 19 | 16681.46 |
| 452 | 2 | 19 | 0 | 20 | 16680.21 |
| 453 | 2 | 20 | 0 | 21 | 16678.90 |
| 454 | 2 | 21 | 0 | 22 | 16677.48 |
| 455 | 2 | 22 | 0 | 23 | 16675.98 |
| 456 | 2 | 23 | 0 | 24 | 16674.39 |
| 457 | 2 | 24 | 0 | 25 | 16672.70 |
| 458 | 2 | 7 | 0 | 6 | 16684.49 |
| 459 | 2 | 8 | 0 | 7 | 16683.49 |
| 460 | 2 | 9 | 0 | 8 | 16682.38 |
| 461 | 2 | 10 | 0 | 9 | 16681.21 |
| 462 | 2 | 11 | 0 | 10 | 16679.96 |
| 463 | 2 | 12 | 0 | 11 | 16678.65 |
| 464 | 2 | 13 | 0 | 12 | 16677.23 |
| 465 | 2 | 14 | 0 | 13 | 16675.70 |
| 466 | 2 | 15 | 0 | 14 | 16674.14 |
| 467 | 2 | 16 | 0 | 15 | 16672.45 |
| 468 | 0 | 7 | 1 | 8 | 17189.13 |
| 469 | 0 | 8 | 1 | 9 | 17188.57 |
| 470 | 0 | 9 | 1 | 10 | 17187.92 |
| 471 | 0 | 10 | 1 | 11 | 17187.21 |
| 472 | 0 | 11 | 1 | 12 | 17186.41 |
| 473 | 0 | 12 | 1 | 13 | 17185.56 |
| 474 | 0 | 13 | 1 | 14 | 17184.61 |
| 475 | 0 | 14 | 1 | 15 | 17183.52 |
| 476 | 0 | 15 | 1 | 16 | 17182.37 |
| 477 | 0 | 17 | 1 | 18 | 17179.77 |
| 478 | 0 | 18 | 1 | 19 | 17178.34 |
| 479 | 0 | 19 | 1 | 20 | 17176.79 |
| 480 | 0 | 20 | 1 | 21 | 17175.18 |
| 481 | 0 | 21 | 1 | 22 | 17173.43 |
| 482 | 0 | 22 | 1 | 23 | 17171.60 |
| 483 | 0 | 24 | 1 | 25 | 17167.60 |
| 484 | 0 | 25 | 1 | 26 | 17165.47 |
| 485 | 0 | 26 | 1 | 27 | 17163.23 |
| 486 | 0 | 9 | 1 | 8 | 17181.81 |
| 487 | 0 | 10 | 1 | 9 | 17180.36 |
| 488 | 0 | 11 | 1 | 10 | 17178.85 |
| 489 | 0 | 12 | 1 | 11 | 17177.29 |
| 490 | 0 | 16 | 1 | 15 | 17170.18 |
| 491 | 0 | 17 | 1 | 16 | 17168.18 |
| 492 | 0 | 21 | 1 | 20 | 17159.19 |
| 493 | 0 | 22 | 1 | 21 | 17156.69 |
| 494 | 0 | 26 | 1 | 25 | 17145.75 |
| 495 | 2 | 8 | 1 | 9 | 16805.09 |
| 496 | 2 | 9 | 1 | 10 | 16804.46 |
| 497 | 2 | 10 | 1 | 11 | 16803.84 |
| 498 | 2 | 11 | 1 | 12 | 16803.11 |
| 499 | 2 | 12 | 1 | 13 | 16802.32 |
| 500 | 2 | 13 | 1 | 14 | 16801.44 |
| 501 | 2 | 14 | 1 | 15 | 16800.48 |
| 502 | 2 | 15 | 1 | 16 | 16799.41 |
| 503 | 2 | 16 | 1 | 17 | 16798.28 |
| 504 | 2 | 17 | 1 | 18 | 16797.04 |
| 505 | 2 | 18 | 1 | 19 | 16795.71 |
| 506 | 2 | 7 | 1 | 6 | 16800.93 |
| 507 | 2 | 8 | 1 | 7 | 16799.66 |
| 508 | 2 | 9 | 1 | 8 | 16798.37 |
| 509 | 2 | 10 | 1 | 9 | 16797.01 |
| 510 | 2 | 11 | 1 | 10 | 16795.57 |
| 511 | 2 | 12 | 1 | 11 | 16794.08 |
| 512 | | | | | |
| 513 | | | | | |
| 514 | 1 | 5 | 1 | 6 | 16997.12 |
| 515 | 1 | 6 | 1 | 7 | 16996.72 |
| 516 | 1 | 7 | 1 | 8 | 16996.22 |
| 517 | 1 | 8 | 1 | 9 | 16995.68 |
| 518 | 1 | 9 | 1 | 10 | 16995.08 |
| 519 | 1 | 10 | 1 | 11 | 16994.40 |
| 520 | 1 | 11 | 1 | 12 | 16993.65 |
| 521 | 1 | 12 | 1 | 13 | 16992.82 |
| 522 | 1 | 13 | 1 | 14 | 16991.89 |
| 523 | 1 | 14 | 1 | 15 | 16990.88 |
| 524 | 1 | 15 | 1 | 16 | 16989.77 |
| 525 | 1 | 16 | 1 | 17 | 16988.57 |
| 526 | 1 | 17 | 1 | 18 | 16987.27 |
| 527 | | | | | |
| 528 | 1 | 9 | 1 | 8 | 16988.98 |
| 529 | 1 | 10 | 1 | 9 | 16987.53 |
| 530 | 1 | 11 | 1 | 10 | 16986.09 |
| 531 | 1 | 12 | 1 | 11 | 16984.56 |
| 532 | 1 | 13 | 1 | 12 | 16982.95 |
| 533 | 1 | 14 | 1 | 13 | 16981.24 |
| 534 | 1 | 15 | 1 | 14 | 16979.51 |
| 535 | 1 | 16 | 1 | 15 | 16977.64 |
| 536 | 1 | 17 | 1 | 16 | 16975.68 |
| 537 | 0 | 4  | 2 | 5  | 17297.54 |
| 538 | 0 | 5  | 2 | 6  | 17297.33 |
| 539 | 0 | 6  | 2 | 7  | 17297.03 |
| 540 | 0 | 7  | 2 | 8  | 17296.58 |
| 541 | 0 | 8  | 2 | 9  | 17296.04 |
| 542 | 0 | 9  | 2 | 10 | 17295.41 |
| 543 | 0 | 10 | 2 | 11 | 17294.61 |
| 544 | 0 | 11 | 2 | 12 | 17293.68 |
| 545 | 0 | 12 | 2 | 13 | 17292.63 |
| 546 | 0 | 13 | 2 | 14 | 17291.44 |
| 547 | 0 | 14 | 2 | 15 | 17290.09 |
| 548 | 0 | 15 | 2 | 16 | 17288.30 |
| 549 | 0 | 16 | 2 | 17 | 17288.54 |
| 550 | 0 | 17 | 2 | 18 | 17286.92 |
| 551 | 0 | 18 | 2 | 19 | 17285.25 |
| 552 | 0 | 19 | 2 | 20 | 17283.49 |
| 553 | 0 | 20 | 2 | 21 | 17281.60 |
| 554 | 0 | 21 | 2 | 22 | 17279.63 |
| 555 | 0 | 22 | 2 | 23 | 17277.57 |
| 556 | 0 | 23 | 2 | 24 | 17275.39 |
| 557 | 0 | 24 | 2 | 25 | 17273.13 |
| 558 | 0 | 6  | 2 | 5  | 17292.81 |
| 559 | 0 | 7  | 2 | 6  | 17291.73 |
| 560 | 0 | 8  | 2 | 7  | 17290.54 |
| 561 | 0 | 9  | 2 | 8  | 17289.25 |
| 562 | 0 | 10 | 2 | 9  | 17287.85 |
| 563 | 0 | 11 | 2 | 10 | 17286.34 |
| 564 | 0 | 12 | 2 | 11 | 17284.68 |
| 565 | 0 | 13 | 2 | 12 | 17282.92 |
| 566 | 0 | 14 | 2 | 13 | 17281.04 |
| 567 | 0 | 15 | 2 | 14 | 17278.95 |
| 568 | 0 | 16 | 2 | 15 | 17276.68 |
| 569 | 0 | 17 | 2 | 16 | 17274.11 |
| 570 | 0 | 18 | 2 | 17 | 17273.45 |
| 571 | 0 | 19 | 2 | 18 | 17271.01 |
| 572 | 0 | 20 | 2 | 19 | 17268.47 |
| 573 | 0 | 21 | 2 | 20 | 17265.85 |
| 574 | 0 | 22 | 2 | 21 | 17263.14 |
| 575 | 0 | 23 | 2 | 22 | 17260.32 |
| 576 | 0 | 24 | 2 | 23 | 17257.42 |
| 577 | 0 | 25 | 2 | 24 | 17254.38 |
| 578 | 0 | 26 | 2 | 25 | 17251.25 |
| 579 | 1 | 3  | 2 | 4  | 17104.69 |
| 580 | 1 | 4  | 2 | 5  | 17104.60 |
| 581 | 1 | 5  | 2 | 6  | 17104.43 |
| 582 | 1 | 6  | 2 | 7  | 17104.13 |
| 583 | 1 | 7  | 2 | 8  | 17103.69 |
| 584 | 1 | 8  | 2 | 9  | 17103.17 |
| 585 | 1 | 9  | 2 | 10 | 17102.55 |
| 586 | 1 | 10 | 2 | 11 | 17101.79 |
| 587 | 1 | 11 | 2 | 12 | 17100.95 |
| 588 | 1 | 12 | 2 | 13 | 17099.92 |
| 589 | 1 | 13 | 2 | 14 | 17098.75 |
| 590 | 1 | 14 | 2 | 15 | 17097.47 |
| 591 | 1 | 15 | 2 | 16 | 17095.68 |
| 592 | 1 | 16 | 2 | 17 | 17095.95 |
| 593 | 1 | 17 | 2 | 18 | 17094.40 |
| 594 | 1 | 18 | 2 | 19 | 17092.80 |
| 595 | 1 | 19 | 2 | 20 | 17091.08 |
| 596 | 1 | 20 | 2 | 21 | 17089.29 |
| 597 | 1 | 21 | 2 | 22 | 17087.38 |
| 598 | 1 | 22 | 2 | 23 | 17085.38 |
| 599 | 1 | 23 | 2 | 24 | 17083.28 |
| 600 | 1 | 6 | 2 | 5 | 17099.89 |
| 601 | 1 | 7 | 2 | 6 | 17098.84 |
| 602 | 1 | 8 | 2 | 7 | 17097.70 |
| 603 | 1 | 9 | 2 | 8 | 17096.41 |
| 604 | 1 | 10 | 2 | 9 | 17095.05 |
| 605 | 1 | 11 | 2 | 10 | 17093.57 |
| 606 | 1 | 12 | 2 | 11 | 17091.96 |
| 607 | 1 | 13 | 2 | 12 | 17090.24 |
| 608 | 1 | 14 | 2 | 13 | 17088.38 |
| 609 | 1 | 15 | 2 | 14 | 17086.34 |
| 610 | 2 | 3 | 2 | 4 | 16914.00 |
| 611 | 2 | 4 | 2 | 5 | 16913.93 |
| 612 | 2 | 5 | 2 | 6 | 16913.74 |
| 613 | 2 | 6 | 2 | 7 | 16913.46 |
| 614 | 2 | 8 | 2 | 9 | 16912.57 |
| 615 | 2 | 9 | 2 | 10 | 16911.97 |
| 616 | 2 | 10 | 2 | 11 | 16911.25 |
| 617 | 2 | 11 | 2 | 12 | 16910.40 |
| 618 | 2 | 12 | 2 | 13 | 16909.44 |
| 619 | 2 | 13 | 2 | 14 | 16908.30 |
| 620 | 2 | 16 | 2 | 17 | 16905.65 |
| 621 | 2 | 17 | 2 | 18 | 16904.16 |
| 622 | 2 | 18 | 2 | 19 | 16902.59 |
| 623 | 2 | 19 | 2 | 20 | 16900.94 |
| 624 | 2 | 22 | 2 | 23 | 16895.44 |
| 625 | 2 | 23 | 2 | 24 | 16893.40 |
| 626 | 2 | 24 | 2 | 25 | 16891.26 |
| 627 | 2 | 27 | 2 | 28 | 16884.23 |
| 628 | 2 | 28 | 2 | 29 | 16881.66 |
| 629 | 2 | 4 | 2 | 3 | 16911.01 |
| 630 | 2 | 5 | 2 | 4 | 16910.18 |
| 631 | 2 | 8 | 2 | 7 | 16907.08 |
| 632 | 2 | 10 | 2 | 9 | 16904.49 |
| 633 | 2 | 11 | 2 | 10 | 16903.04 |
| 634 | 2 | 12 | 2 | 11 | 16901.47 |
| 635 | 2 | 15 | 2 | 14 | 16895.99 |
| 636 | 2 | 16 | 2 | 15 | 16893.80 |
| 637 | 2 | 17 | 2 | 16 | 16891.35 |
| 638 | 2 | 18 | 2 | 17 | 16890.82 |
| 639 | 2 | 21 | 2 | 20 | 16883.58 |
| 640 | 0 | 5 | 3 | 6 | 17398.96 |
| 641 | 0 | 6 | 3 | 7 | 17398.57 |
| 642 | 0 | 8 | 3 | 9 | 17397.39 |
| 643 | 0 | 9 | 3 | 10 | 17396.69 |
| 644 | 0 | 10 | 3 | 11 | 17395.84 |
| 645 | 0 | 11 | 3 | 12 | 17394.87 |
| 646 | 0 | 12 | 3 | 13 | 17393.79 |
| 647 | 0 | 13 | 3 | 14 | 17392.58 |
| 648 | 0 | 14 | 3 | 15 | 17391.24 |
| 649 | 0 | 15 | 3 | 16 | 17389.79 |
| 650 | 0 | 16 | 3 | 17 | 17388.24 |
| 651 | 0 | 17 | 3 | 18 | 17386.56 |
| 652 | 0 | 18 | 3 | 19 | 17384.73 |
| 653 | 0 | 19 | 3 | 20 | 17382.79 |
| 654 | 0 | 9 | 3 | 8 | 17390.70 |
| 655 | 0 | 10 | 3 | 9 | 17389.22 |
| 656 | 0 | 11 | 3 | 10 | 17387.65 |
| 657 | 0 | 12 | 3 | 11 | 17385.92 |
| 658 | 0 | 13 | 3 | 12 | 17384.08 |
| 659 | 0 | 14 | 3 | 13 | 17382.15 |
| 660 | 0 | 15 | 3 | 14 | 17380.08 |
| 661 | 0 | 16 | 3 | 15 | 17377.89 |
| 662 | 0 | 17 | 3 | 16 | 17375.59 |
| 663 | 0 | 18 | 3 | 17 | 17373.18 |
| 664 | 0 | 19 | 3 | 18 | 17370.61 |
| 665 | 0 | 4 | 4 | 5 | 17494.02 |
| 666 | 0 | 5 | 4 | 6 | 17493.68 |
| 667 | 0 | 6 | 4 | 7 | 17493.22 |
| 668 | 0 | 7 | 4 | 8 | 17492.64 |
| 669 | 0 | 8 | 4 | 9 | 17491.92 |
| 670 | 0 | 9 | 4 | 10 | 17491.08 |
| 671 | 0 | 10 | 4 | 11 | 17490.13 |
| 672 | 0 | 11 | 4 | 12 | 17489.03 |
| 673 | 0 | 3 | 5 | 4 | 17582.48 |
| 674 | 0 | 4 | 5 | 5 | 17582.23 |
| 675 | 0 | 5 | 5 | 6 | 17581.86 |
| 676 | 0 | 6 | 5 | 7 | 17581.32 |
| 677 | 0 | 7 | 5 | 8 | 17580.64 |
| 678 | 0 | 8 | 5 | 9 | 17579.82 |
| 679 | 0 | 9 | 5 | 10 | 17578.89 |
| 680 | 0 | 10 | 5 | 11 | 17577.81 |